# A Fast-rate WLAN Measurement Tool for Improved Miss-rate in Indoor Navigation

Erick Schmidt, *The University of Texas at San Antonio*
David Akopian, *The University of Texas at San Antonio*

## BIOGRAPHIES

**Erick Schmidt** received his B.S. degree from Monterrey Institute of Technology and Higher Education, Mexico, in 2011 and his M.S. degree from The University of Texas at San Antonio (UTSA), in 2015. He is currently pursuing his Ph.D. degree at UTSA. His research interests include software-defined radio for fast prototyping, WLAN indoor localization, and interference mitigation techniques for Global Navigation Satellite System (GNSS).

**David Akopian** is a Professor at the University of Texas at San Antonio (UTSA). Dr. Akopian received his Ph.D. degree from Tampere University of Technology, Finland, in 1997. Dr. Akopian's current research interests include signal processing algorithms for communication and navigation receivers, and implementation platforms for software-defined radio, and mHealth. Dr. Akopian is a Senior Member of IEEE, member of the Institute of Navigation (ION), and Fellow of National Academy of Inventors (NAI).

## ABSTRACT

Recently, location-based services (LBS) have steered attention to indoor positioning systems (IPS). WLAN-based IPSs relying on received signal strength (RSS) measurements such as fingerprinting are gaining popularity due to proven high accuracy of their results. Typically, sets of RSS measurements at selected locations from several WLAN access points (APs) are used to calibrate the system. Retrieval of such measurements from WLAN cards are commonly at one-Hz rate. Such measurement collection is needed for offline radio-map surveying stage which aligns fingerprints to locations, and for online navigation stage, when collected measurements are associated with the radio-map for user navigation. As WLAN network is not originally designed for positioning, an RSS measurement miss could have a high impact on the fingerprinting system. Additionally, measurement fluctuations require laborious signal processing, and surveying process can be very time consuming. This paper proposes a fast-rate measurement collection method that addresses previously mentioned problems by achieving a higher probability of RSS measurement collection during a given one-second window. This translates to more data for statistical processing and faster surveying. The fast-rate collection approach is analyzed against the conventional measurement rate in a proposed testing methodology that mimics real-life scenarios related to IPS surveying and online navigation.

## 1 INTRODUCTION

The prospect of indoor navigation has gained attention in the last decade based on ubiquitous applications in location-based services (LBS) such as commercial, emergency, and military. While outdoor navigation is achieved by existing global navigation satellite systems (GNSS) such as the Global Positioning System (GPS) [1], indoor navigation is inaccessible for GPS signals due to structural blockages and severe multipath propagation effects. Currently, there is no general solution for indoor navigation but there are proposed techniques that use measurements from deployed indoor wireless signaling infrastructures such as Wi-Fi and Wireless Local Area Networks (WLANs) [2]. Several suggested solutions based on WLAN signals and conventional positioning techniques such as time-of-arrival (TOA) [3], time-difference-of-arrival (TDOA) [4], [5], ranging-trilateration [6], [7], and angle-of-arrival (AOA) [8], [9], can be found in literature. Specifically, many WLAN-based indoor positioning systems (IPS) are relying on received signal strength (RSS) indicators for localization, such as fingerprinting-based [10], and propagation modeling-based [11], among others. Still, there are many challenges related to WLAN-based positioning that should be addressed for a broader and more robust deployment.

The fingerprinting-based IPS approach is among the most popular due to its high-accuracy results as seen in [2], [10]. As opposed to analytical positioning methods such as AOA, TOA, etc., which rely on bounded models, fingerprinting-based methods actually leverage on complex and cluttered indoor settings. The fingerprinting solution is based on an empirical method which collects RSS measurements as "signatures" (or fingerprints) for known grid locations and designs a database which associates the collected fingerprints with said grid locations, known as a "radio-map". The empirical technique thus has two stages of operation: offline and online mode. The offline stage collects RSS measurements for each discrete location to calibrate the radio-map. Afterwards, the online stage processes user captured RSS measurements followed by a matching of said fingerprint to the closest radio-map entry [10]. Depending on the mode of matching estimation, these classification methods are categorized as deterministic [2], probabilistic [10], machine learning [12], [13], among others.

Many solutions in literature typically focus on the online stage of said IPS, specifically on location and classification techniques, assuming reliable pre-collected offline radio-maps. Nonetheless, the offline stage of radio-map measurement collection is as equally important as the online navigation stage for a robust operation [14], [15]. Typically, these indoor surveying and navigation stages are manpower and time consuming, while at the same time the performance of radio-map construction is critical [15]. Since WLAN infrastructure was not built for indoor navigation, said collected fingerprint measurements might not always guarantee a reliable performance. Therefore, several phenomena have been noted in both indoor offline and online stages of IPS, related mostly to the reliability and availability of RSS measurements. One proposed solution to the reliability problem is the selection of RSS measurements from stable WLAN access points (APs) [16]. Another approach to improve the reliability of the measurements is by detecting and filtering "outliers" from the available measurement set [17]-[19]. Distortion impact of various outlier types on WLAN-based fingerprint positioning as well as missed measurement effects is reported in [20], [21]. Therefore, the availability and reliability of measurements during offline surveying as well as online navigation is important.

When capturing RSS measurements from mobile devices, data collecting tools such as Wireshark often see measurement rate restrictions due to numerous factors such as WLAN card vendor limitations, the card driver itself, operating system (OS) compatibility, among others [22]. A well-known one-Hz measurement collection rate is commonly reported in literature [14], [23], [24], which increases the importance of measurement availability per each second period for both offline and online stages. Avoiding a missed measurement in said second period becomes significant. Additionally, advanced probabilistic and machine-learning techniques require hundreds of measurements at each location for statistically sound results for calibration in offline as well as navigation in online stages, respectively. Said missed measurements (or outliers) that hinder IPS performance have been reported to occur mainly due to three causes: (1) faulty APs, (2) WLAN interference, and (3) transient effects on WLAN cards [25]. A first attempt to address this missed measurement issue has been reported in [26] by applying interpolation techniques or decimating the location grid map, but naturally results in degraded accuracy. Additionally, advanced filtering algorithms have been proposed for outlier detection including missing measurements caused by transient effects in network cards or unwanted interference, but their performance typically depends on the number of available APs [19], [26], and [27].

The previously mentioned issues related to a missed measurement or packet on the given one second period are addressed by proposing a method of capturing RSS packets from a single AP at about ten times faster than the one-Hz rate an open-source tool [25]. This is followed by an assessment of measurement rate and measurement availability by presenting a methodology for testing said fast-rate measurement extraction mode vs. conventional one-Hz methods. A higher rate measurement capture is expected to increase measurement availability within the common one second analysis period. Additionally, it accelerates proportionally radio-map surveying time through faster sample collection and aids in calibration of highly demanding probabilistic and machine learning techniques due to the same reason. In a one second period, the probability of availability of an RSS sample is drastically increased with the proposed method, thus almost nullifying outliers, specifically missed packets which degrade the performance of WLAN positioning [20], [21], [23], [26], and [27]. Therefore, the study presented in this paper focuses mostly on missing measurements resulting from previously mentioned outlier phenomena: (1) faulty APs, (2) unintentional WLAN interference, and (3) possible WLAN card transient effects [25].

## 2 MEASUREMENT COLLECTION MODES

WLAN's main functionality as a wireless communication medium tends to have several sources of traffic in terms of the interaction between several APs and mobile devices. As an attempt to minimize traffic for a specific mobile device, the WLAN cards have built-in capturing filters. There are two filters related to this dedicated connection between APs and mobile devices: WLAN channel filter which focuses the connection to one of 11 radio channels listed on the 802.11 standard [28], and service set identifier (SSID) filter which relates to the current connected network name and ignores most of traffic from other SSID networks. When using a capturing tool for RSS measurements, these filters unintentionally impede capturing rate as Wi-Fi networks are not built for IPS, especially when the mobile device has a dedicated connection.

Figure 1 shows the interaction of RSS measurements coming from WLAN packets with capturing filters inside the WLAN card. The capturing tool would then capture in so-called "normal mode", thus limiting the capturing rate. For this mode, the capturing steps can be seen as follows: WLAN card → capturing filter → capturing tool → hard drive. The majority of the WLAN-fingerprint measurement collection solutions in literature exploit this type of data collection [25]. To overcome this limitations, this paper proposes bypassing previously mentioned capturing filters by operating the WLAN card in so-called "monitor mode". In monitor mode, WLAN packets for all 11 WLAN channels and all detectable SSID names for APs are visible for the capturing tool. This can be accomplished by, e.g. Aircrack [29], an open-source tool that modifies operation mode of the WLAN card to monitor mode by modifying the WLAN card driver. The WLAN card driver works as an interface between the card, the OS, and the capturing tool found in the host PC. The capturing steps are now: WLAN card → capturing tool → hard drive (see Figure 1). Therefore, this paper examines both normal and monitor modes for measurement collection. An overview of the physical (PHY) layer of the WLAN standard is shown next.

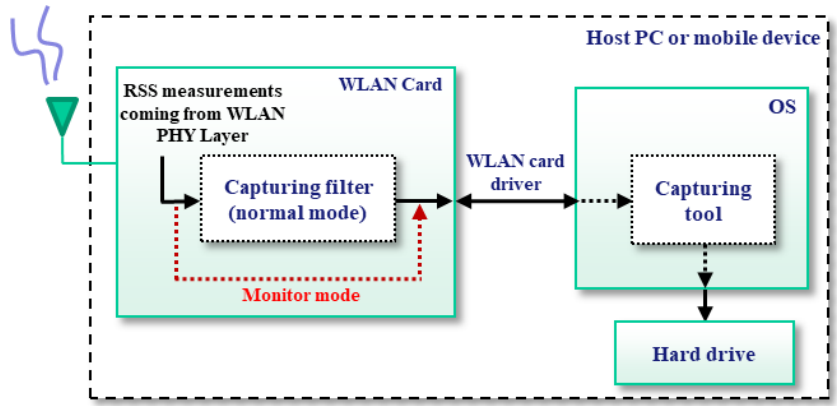

**Figure 1. Interface diagram showing the interaction for both normal and monitor RSS collection modes.**

### 2.1 WLAN beacon frames

To capture RSS measurements, the WLAN card requires specific broadcast packets that are coming from all APs and contain other information to relate to the networks being broadcasted, such as the SSID, and the medium access control (MAC) address, among others. These packets are called "beacon frames". After capturing said packets, the WLAN card measures the RSS internally at the end of the packet. The structure of a PHY packet data unit (PPDU) can be seen in Figure 2. It is composed of a PHY and a MAC portions. The PHY layer contains a known preamble used to detect the presence of a packet as well as a header with general information used to demodulate the rest of the MAC portion, namely the MAC packet data unit (MPDU). On the MPDU, there is a MAC header with specific information such as the packet type and subtype, which for the purposes of RSS measurement capture, should be management type and beacon frame subtype. On the Frame Body of the MPDU, other relevant information such as the SSID, can be found. Technically speaking, the WLAN card will report RSS every time it successfully decodes the full beacon frame.

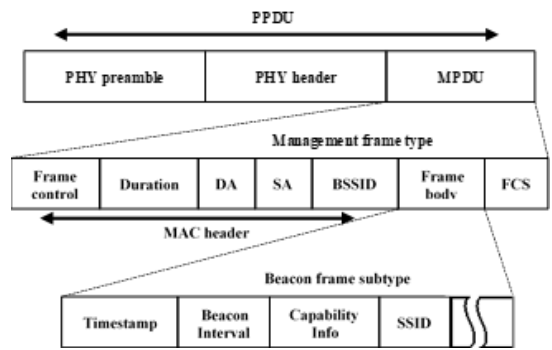

**Figure 2. WLAN physical layer packet data unit structure for beacon frames.**

**2.2 Beacon intervals**

The 802.11 standard uses a time slot system with a defined time unit (TU) corresponding to 1024 microseconds. As per the standard, the fastest interval for a beacon frame transmission is every 100 TUs, or 102.4 milliseconds. Therefore, assuming an unobstructed measurement collection from a WLAN card, we can theoretically acquire around 9.77 RSS packets per second (pps). The "monitor mode" is used for this purpose. On the other hand, normal mode experiments demonstrate measurement rate interval of 1,000 TUs which is every 1024 milliseconds (roughly a second). This translates to theoretical rate of 0.977 pps for normal mode. These theoretical maximum values will be used in our tests: 9.77 pps for monitor mode, and 0.977 pps for normal mode, respectively.

**3 TESTING METHODOLOGY**

We present a testing methodology to assess representative surveying situations in both monitor and normal capturing modes. With three specific scenarios, we attempt to observe capturing capabilities and measurement data collection rate differences for these two modes of operation, while at the same time examine previously mentioned measurement effects: faulty APs, interference, and transient effects. The tests are: traffic test, distance (signal strength) test, and vendor CPU load test. Said tests are performed in representative environments at the University of Texas at San Antonio (UTSA), and a nearby student apartment complex.

**3.1 Testing specifications**

At the moment of this writing, Aircrack measurement tool has sole compatibility with Linux OS [29]. Therefore, the three scenarios in normal mode are done in both Windows and Linux OS, while monitor mode is done in Linux only. The equipment used was an ASUS X555LA series laptop with a 4th generation Intel Core i5 at 1.7 GHz, 8GB of RAM, and Windows 10 Home, along with Linux Ubuntu 16.04 LTS in a dual-boot mode. As for the WLAN cards, an integrated Atheros model AR9485 with 802.11b/g/n capabilities, and a Ralink RT3070 USB dongle with similar Wi-Fi capabilities as a second device were used. Both cards are compatible with Aircrack in Linux. The capturing tools were Wireshark version 2.0.12 for Linux [22], and Acrylic Wi-Fi Professional v3.2.6269.20454 [30] for Windows. For a statistically sound analysis, a total of 200 seconds capturing time, performed 10 times, was performed for each testing scenario. This gives a total of 33.33 minutes worth of capturing time per scenario. The testing scenarios are described next.

**3.2 Traffic test**

This test is built upon an interference hypothesis in which high WLAN traffic will negatively impact measurement collection rates. Therefore, a congested environment is elaborated for this testing scenario. The traffic test was performed in an apartment near the university, where 4 WLAN routers were placed throughout the apartment and three reference points (RPs) were used for measurement collection. Additionally, 5 people were present at the apartment during said test. Figure 3 shows the floor plan and testing configuration for said APs and RPs.

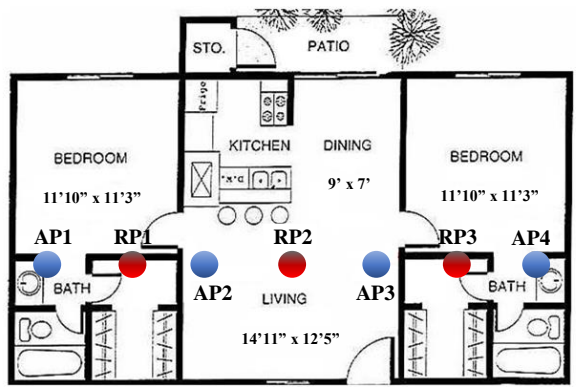

**Figure 3. Apartment floorplan used in traffic test.**

### 3.3 Distance (signal strength) test

This test assesses the impact of distance of the mobile device and the AP in terms of RSS. The hypothesis is that said distance variation might degrade capturing rate of the WLAN card and thus produce misses due to faulty APs. Weaker signal conditions (or further distances) might produce a packet miss for the given one-second period measurement window in normal mode [25]. An improvements in measurement availability due to monitor mode is expected. One AP was placed in a hallway of the applied engineering and technology (AET) building at the university. Two distances with averaged RSS values of -60 dBm and -80 dBm were measured for strong and weak signal levels, respectively. This test was performed on Atheros NIC for normal and monitor modes. Figure 4 shows the floorplan configuration.

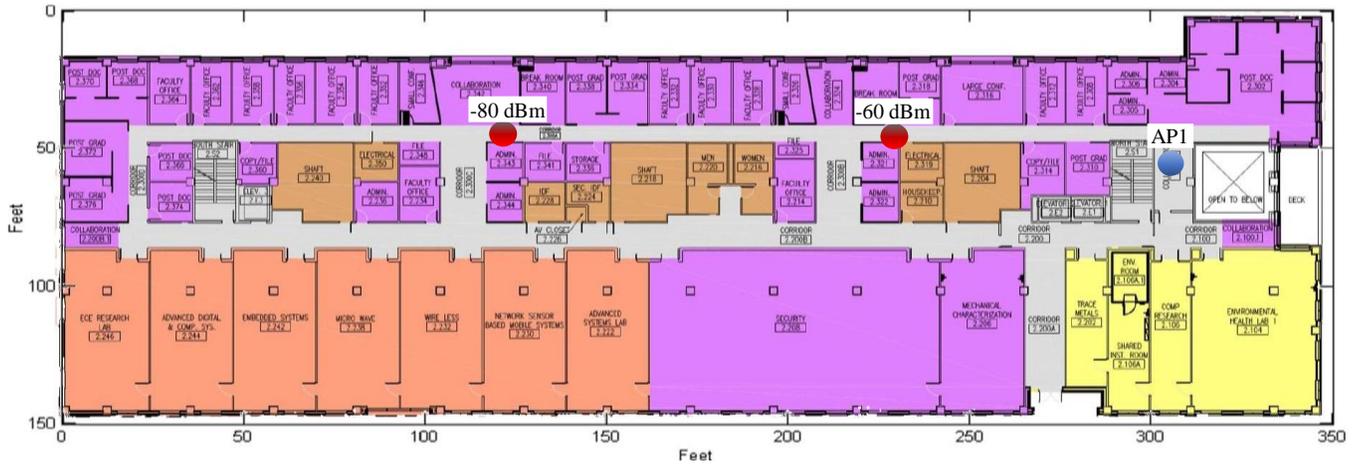

**Figure 4. UTSA's AET building hallway for distance (signal strength) and vendor CPU load tests.**

### 3.4 Vendor CPU load test

In this test, an artificial CPU load on the mobile capturing device is assessed. It is hypothesized that this test can results in transient effects on the WLAN card, thus evidencing missed packet effects on the device. For this test, a CPU stress tool Heavy Load v3.4 [31] is used in Windows, and stress-ng [32] is used in Linux. Test scenarios of 50% and 80% CPU load were selected, for both normal mode and monitor mode, for both Atheros and Ralink NICs. This to see the transient effects on different loads and different card vendors, for both capturing modes. The test was performed in weak signal conditions (-80dBm) to compare against a probable worst-case situation for either offline surveying or online navigation. This test was performed in the same hallway location as the distance test (see Figure 4).

### 4 RESULTS

For all three scenarios, two statistics were used: average measure rate, in pps, and miss-rate compared to theoretical maximum, in percentage. For the theoretical maximums, 9.77 pps and 0.97 pps were used for monitor mode and normal mode respectively. Additionally, the measurement timestamps were evaluated against interval histograms based on the time difference of packet arrival between consecutive packets. With this time intervals of around 100 ms, a probability density function (PDF) and cumulative distribution function (CDF) plot was created based on histogram of different arrival time intervals.

### 4.1 Traffic test results

Table 1 shows the results for traffic test for both capturing modes. The results for all three RPs were averaged to show a single result for each AP. It can be seen that normal mode is less affected in high traffic environments. The WLAN card capturing filter ignores most of the traffic in the environment, therefore achieving a low packet loss of 0.90 pps and around 7% of miss-rate compared to the theoretical maximum value. Conversely, since monitor mode captures unrestricted traffic, a possible interference occurs resulting in miss-rate of up to 51% and 4.76 pps. Nonetheless, worst-case monitor mode still captures more measurement per second (4.76 pps) compared to best-case normal mode (0.91 pps).

**Table 1. Traffic test results for normal and monitor modes.**

| Mode | Normal mode | | Monitor mode | |
|---|---|---|---|---|
| | Avg. meas. rate. (pps) | Miss-rate (%) * | Avg. meas. rate. (pps) | Miss-rate (%) * |
| AP1 | 0.90 | 7.50% | **4.76** | **51.25%** |
| AP2 | 0.91 | 7.09% | 7.40 | 24.22% |
| AP3 | 0.91 | 6.93% | 7.42 | 24.06% |
| AP4 | 0.91 | 7.03% | 8.29 | 15.10% |

*Miss-rate compared to theoretical maximum for each capturing mode (see Section 2.2)

## 4.2 Distance (signal strength) test results

Table 2 shows distance test results for both strong and weak signal strengths or distances, for normal and monitor modes. Normal mode showed a noticeable impact when increasing AP distance from mobile device, by showing a miss-rate of 42% and 0.57 pps which translates to almost half a packet per second. It's hypothesized that this is due to the capturing filters in normal mode. Monitor mode shows barely any impact as its unfiltered capturing is actively listening to even far distance packets. The miss-rate for the near distance is close to the theoretical maximum, and for the far distance, a small degradation of around 5% miss-rate was seen, still showing a fast-rate of 9.22 pps nonetheless.

**Table 2. Distance (signal strength) test results for normal and monitor modes.**

| Mode | Normal mode | | Monitor mode | |
|---|---|---|---|---|
| | Avg. meas. rate, packets per sec. | Miss-rate (%) * | Avg. meas. rate, packets per sec. | Miss-rate (%) * |
| -60 dBm (strong/near) | 0.91 | 7.04% | 9.68 | 0.86% |
| -80 dBm (weak/far) | **0.57** | **41.67%** | 9.22 | 5.63% |

Figure 5 shows the PDF and CDF plot to better depict this packet loss phenomena due to faulty APs for normal mode when distance is increased on the mobile device. A clear evidence of packet arrival near the 2000 ms interval (or 2 second interval) can be seen when at -80 dBm. This translates to a packet miss in the one-second window interval and can highly impact measurement availability. In addition, this observed phenomena for the distance test led to the vendor CPU load test to compare this same weak-signal condition to an induced CPU load. This is to determine transient effects on WLAN cards, while capturing packets on a realistic environment of a potential multitasking mobile device while surveying or navigating.

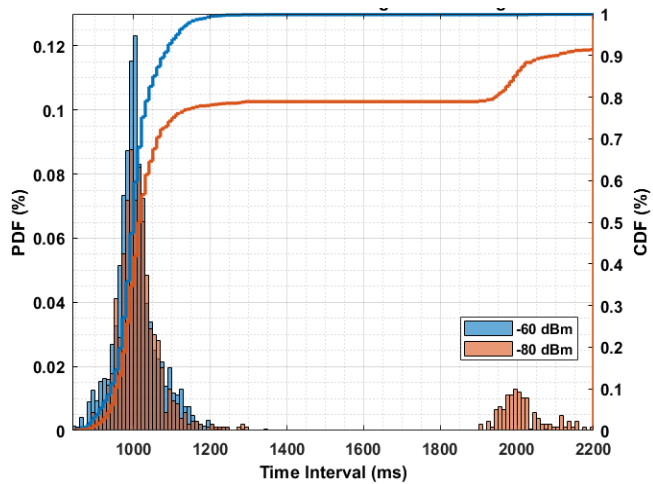

**Figure 5. Distance test time interval histogram in normal mode for strong/near conditions (-60 dBm) and weak signal conditions (-80 dBm).**

### 4.3 Vendor CPU load test results

As seen in the distance (signal strength) test results section, an impact on measurement rate is expected in this test based on the added stress on the CPU. Table 3 shows vendor comparison CPU load tests (at weak-signal distance) for normal mode. Overall, Ralink underperformed in measurement rate compared to Atheros while both experienced a degradation of roughly 11% when CPU stress is applied. Ralink performed worse with a miss-rate of nearly 30% and 0.7 pps.

**Table 3. Vendor CPU load test results for normal mode for Atheros and Ralink vendors.**

| Mode | Atheros (integrated) | | Ralink (USB dongle) | |
|---|---|---|---|---|
| | Avg. meas. rate, packets per sec. | Miss-rate (%) * | Avg. meas. rate, packets per sec. | Miss-rate (%) * |
| **50% load** | 0.96 | 2.17% | 0.76 | 21.92% |
| **80% load** | 0.86 | 11.86% | **0.69** | **29.45%** |

To demonstrate transient effects and similarly to the distance test in Section 4.2, Figure 6 shows the histogram chart for vendor CPU test for Atheros in normal mode for both 50% and 80% CPU loads (similar trend is observed in Ralink NIC results). A spread from the 1 second main lobe vicinity towards the 2 second delays can be seen when stress is increased to 80% load on weak signal conditions.

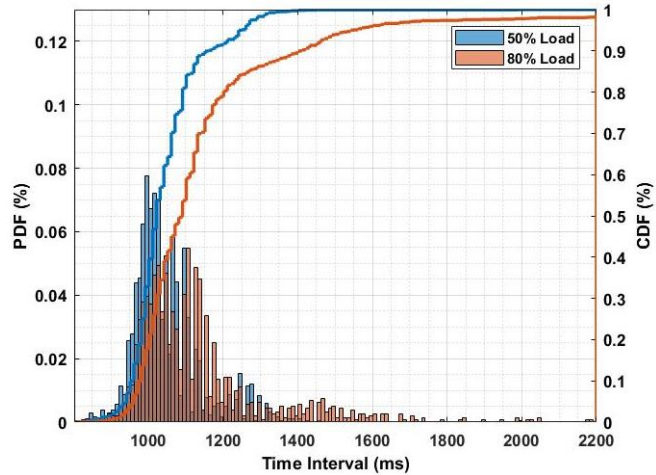

**Figure 6. Atheros vendor CPU load (stress) test time interval histogram in normal mode for 50% and 80% CPU load at weak signal conditions (-80 dBm).**

Table 4 shows vendor CPU load tests for monitor mode for both Atheros and Ralink. Similar to normal mode, there is an evident transient effect in monitor mode when CPU load is increased. In this case, the Ralink vendor showed a more evident transient effect by reaching a miss-rate of 48% when compared to theoretical maximum, and 5 pps. This transient effect from the vendor CPU test occurs similar to what is seen in Figure 6 for normal mode. This 5 pps rate translates to interval miss values around 500 to 600 ms, corresponding to 5 to 6 packet misses from the theoretical maximum interval near the 100ms [25].

**Table 4. Vendor CPU load test results for monitor mode for Atheros and Ralink vendors.**

| Mode | Atheros (integrated) | | Ralink (USB dongle) | |
|---|---|---|---|---|
| | Avg. meas. rate, packets per sec. | Miss-rate (%) * | Avg. meas. rate, packets per sec. | Miss-rate (%) * |
| **50% load** | 8.76 | 10.28% | 8.19 | 16.13% |
| **80% load** | 7.80 | 20.17% | **5.05** | **48.30%** |

## 5 CONCLUSION

This paper proposed a fast-rate capturing method for RSS measurements at about ten times faster than conventional one-Hz capture rates for IPS. Furthermore, it proposed a testing methodology to address a representative performance study of these two capturing modes. The proposed fast-rate method guarantees much higher availability of a RSS measurement on a one second period, therefore minimizing missed measurement phenomena. To achieve this, an open-source tool enables a so-called monitor mode by means of the WLAN card drivers, as opposed to normal mode. As the measurement capture is a random process, the paper assessed a performance study of three scenarios to better depict real situations: traffic test, distance test, and vendor CPU test. For worst-case scenarios, monitor mode suffered a miss-rate of 51% in the traffic test but still achieved a rate of 4.8 pps. On the other hand, the worst-case scenario for normal mode was the distance test which saw a degradation of 42% miss-rate, corresponding to a rate of 0.5 pps. Nonetheless, the proposed testing methodology leaves space for many scenarios as future work. Additionally, the open-source tool constantly adds more WLAN vendors, thus an applicability for mobile devices such as smartphones is also a testing candidate.

## REFERENCES

[1] Misra, P. and Enge, P., *Global Positioning System: Signals, Measurements, and Performance*, 2nd ed. Ganga-Jamuna Press, Lincoln MA, 2006.
[2] Khalajmehrabadi, A., Gatsis, N. and Akopian, D., "Modern WLAN fingerprinting indoor positioning methods and deployment challenges," *IEEE Communication Surveys and Tutorials*, Vol. 19, No. 3, 2017, pp. 1974-2002.
[3] Li, X. and Pahlavan, K., "Super-resolution TOA estimation with diversity for indoor geolocation," *IEEE Transactions on Wireless Communications*, Vol. 3, No. 1, Jan. 2004, pp. 224–234.
[4] Comsa, C.-R., Luo, J., Haimovich, A. and Schwartz, S., "Wireless localization using time difference of arrival in narrow-band multipath systems," in *International Symposium on Signals, Circuits and Systems*, Vol. 2, Jul. 2007, pp. 1–4.
[5] Makki, A., Siddig, A., Saad, M., Cavallaro, J.R. and Bleakley, C.J., "Indoor Localization Using 802.11 Time Differences of Arrival," *IEEE Transactions on Instrumentation and Measurements*, Vol.65, No.3, 2016, pp.614-623.
[6] Rusli, M. E., Ali, M., Jamil, N. and Din, M. M., "An Improved Indoor Positioning Algorithm Based on RSSI-Trilateration technique for Internet of Things (IoT)," in *Computer and Communication Engineering (ICCCE), 2016 International Conference on*, Kuala Lumpur, Malaysia, 2016.
[7] He, S. and Chan, S.-H. G., "INTRI: Contour-Based Trilateration for Indoor Fingerprint-Based Localization," *IEEE Transactions on Mobile Computing*, Vol. 16, No. 6, Jun. 2017, pp. 1676 - 1690.
[8] Biswas, P., Aghajan, H. and Ye, Y., "Integration of angle of arrival information for multimodal sensor network localization using semidefinite programming," in *Proceedings of the 39th Asilomar Conference on Signals, Systems and Computers*, Oct. 2005.
[9] Niculescu, D. and Nath, B., "Ad hoc positioning system (APS) using AOA," in *Proceedings of the 22nd Annual Joint Conference of the IEEE Computer and Communications*, Vol. 3, Mar. 2003, pp. 1734–1743.
[10] He, S. and Chan, S.-H. G., "Wi-Fi Fingerprint-Based Indoor Positioning: Recent Advances and Comparisons," *IEEE Communication Surveys and Tutorials*, Vol. 18, No. 1, 2016, pp. 466-490.
[11] Shen, X., Xu, K., Sun X., Wu, J. and Lin, J., "Optimized indoor wireless propagation model in WiFi-RoF network architecture for RSS-based localization in the Internet of Things," in *Microwave Photonics, 2011 International Topical Meeting on & Microwave Photonics Conference, 2011 Asia-Pacific, MWP/APMP*, Singapore, Singapore, 2011.
[12] Salamah, A. H., Tamazin, M., Sharkas, M. A. and Khedr, M., "An enhanced WiFi indoor localization system based on machine learning," in *Indoor Positioning and Indoor Navigation (IPIN), 2016 International Conference on*, Alcala de Henares, Spain, 2016.
[13] Wang, X., Gao, L., Mao, S. and Pandey, S., "CSI-Based Fingerprinting for Indoor Localization: A Deep Learning Approach," *IEEE Transactions on Vehicular Technology*, Vol. 66, No. 1, 2017, pp. 763-776.
[14] Jung, S. H., Moon, B.-C. and Han, D., "Performance Evaluation of Radio Map Construction Methods for Wi-Fi Positioning Systems," *IEEE Transactions on Intelligent Transportation Systems*, Vol. 18, No. 4, Apr. 2017, pp. 880-889.
[15] Shih, C., Chen, L., Chen, G., Wu, E. H. and Jin, M., "Intelligent radio map management for future WLAN indoor location fingerprinting," in *Wireless Communications and Networking Conference (WCNC), 2012 IEEE*, Shanghai, China, 2012.
[16] Chen, Y., Yang, Q., Yin, J. and Chai, X., "Power-efficient access point selection for indoor location estimation," *IEEE Transactions on Knowledge and Data Engineering*, Vol. 18, No. 7, Jul. 2006, pp. 877–888.
[17] Hodge, V. and Austin, J., "A survey of outlier detection methodologies," *Artificial Intelligence Review*, Vol. 22, No. 2, Oct. 2004, pp. 85–126.
[18] Chen, Y. C., Sun, W. C. and Juang, J. C., "Outlier detection technique for RSS-based localization problems in wireless sensor networks," in *Proceedings of SICE Annual Conference*, Aug. 2010, pp. 657–662.

[19] Khalajmehrabadi, A., Gatsis, N., Pack, D. and Akopian, D., "A joint indoor WLAN localization and outlier detection scheme using LASSO and Elastic-Net optimization techniques," IEEE Transactions on Mobile Computing, Vol. 16, No.8, 2017, pp. 2079 - 2092.
[20] Morales, J. A., Akopian, D. and Agaian, S., "Faulty measurements impact on wireless local area network positioning performance," *IET Radar, Sonar & Navigation*, Vol. 9, No. 5, Aug. 2015, pp. 501–508.
[21] Morales, J. A., Akopian, D. and Agaian, S., "Mitigating anomalous measurements for indoor wireless local area network positioning," *IET Radar, Sonar & Navigation*, Vol. 10, No. 7, 2016, pp. 1220-1227.
[22] Wireshark, "CaptureSetup/WLAN - The Wireshark Wiki," Wireshark, [Online]. Available: https://wiki.wireshark.org/CaptureSetup/WLAN. [Accessed May 2017].
[23] Laoudias, C., Michaelides, M. P., Panayiotou, C. G., "Fault Detection and Mitigation in WLAN RSS Fingerprint-based Positioning," in *Indoor Positioning and Indoor Navigation (IPIN), 2011 International Conference on*, Guimaraes, Portugal, Sept. 2011.
[24] Kushki, A., Plataniotis, K. N., Venetsanopoulos, A. N., "Kernel-Based Positioning in Wireless Local Area Networks," *IEEE Transactions on Mobile Computing*, Vol. 6, No. 6, Jun. 2007, pp. 689-705.
[25] Schmidt, E., Mohammed, M. A. and Akopian, D., "A Performance Study of a Fast-Rate WLAN Fingerprinting Measurement Collection Method," *IEEE Transactions on Instrumentation and Measurements*, Vol. 67, No. 10, Oct. 2018, pp. 2273–2281.
[26] Talvitie, J.., Renfors, M. and Lohan, E. S., "Distance-based interpolation and extrapolation methods for RSS-based localization with indoor wireless signals," *IEEE Transactions on Vehicular Technology*, Vol. 64, No. 4, Apr. 2015, pp. 1340–1353.
[27] Talvitie, J., Lohan, E. S., Renfors, M., "The effect of coverage gaps and measurement inaccuracies in fingerprinting based indoor localization," in *Localization and GNSS (ICL-GNSS), 2014 International Conference on*, Helsinki, Finland, 2014.
[28] IEEE Standard 802.11-2012 - Part 11: Wireless LAN Medium Access Control (MAC) and Physical Layer (PHY) Specifications, IEEE Standards Association, The Institute of Electrical and Electronics Engineers, Inc, 2012.
[29] Aircrack-ng, Aircrack-ng, [Online]. Available: http://aircrack-ng.org/. [Accessed December 2016].
[30] Acrylic WiFi. Free WiFi scanner and channel scanner for windows, [Online]. Available: https://www.acrylicwifi.com/en/wlan-software/wlan-scanner-acrylic-wifi-free/. [Accessed December 2016].
[31] Jam Software, HeavyLoad - Free Stress Tool for Your PC. Jam Software, [Online]. Available: https://www.jam-software.com/heavyload/. [Accessed Decemeber 2016].
[32] Ubuntu Kernel Team, stress-ng, [Online]. Available: http://kernel.ubuntu.com/~cking/stress-ng/. [Accessed December 2016]